\documentclass[prl,twocolumn,final]{revtex4}
\usepackage{graphicx,color}
   \usepackage{amsmath,amssymb,pifont}

\newcommand{\nn}{\nonumber}

\newcommand{\tr}{\mathrm{tr}}
\newcommand{\Tr}{\mathrm{Tr}}

\renewcommand{\(}{\left(}
\renewcommand{\)}{\right)}
\renewcommand{\[}{\left[}
\renewcommand{\]}{\right]}

\begin{document}
\title{Soft-/rapidity- anomalous dimensions correspondence}
\author{Alexey A. Vladimirov}
\affiliation{Institut f\"ur Theoretische Physik, Universit\"at Regensburg,
D-93040 Regensburg, Germany}

\begin{abstract}
We establish a correspondence between ultraviolet singularities of soft factors for multiparticle production and rapidity singularities of soft factors for multiparton scattering. This correspondence is a consequence of the conformal mapping between scattering geometries. The correspondence is valid to all orders of perturbation theory and in this way provides one with a proof of rapidity renormalization procedure for multiparton scattering (including the transverse momentum dependent (TMD) factorization as a special case). As a by-product, we obtain an exact relation between the rapidity anomalous dimension and the well-known soft anomalous dimension. The three-loop expressions for TMD and a general multiparton scattering rapidity anomalous dimension are derived.
\end{abstract}
\maketitle

\textbf{Introduction}. Factorization theorems are an effective tool for the description of hadron reactions within the perturbative Quantum Chromodynamics (pQCD)\cite{Collins:1989gx,Ellis:2009wj,Bauer:2008jx,Feige:2014wja,Contopanagos:1996nh}. Factorization formulas have a common structure which includes a hard part, parton distributions, jet functions, and the soft factor(s). However, the operator structure of these ingredients can differ drastically for different processes, that leads to significant fragmentation of theoretical results. In this Letter, we discuss a correspondence between soft factors (SFs) typical for different kinematics and consequences of this correspondence.

Generally, SFs represent the soft part of the between-parton interaction. A typical SF is given by a vacuum matrix element of a configuration of Wilson lines that reflects the classical picture of scattering. Being in many aspects artificial, SFs contain a set of infrared divergences and are defined only within an appropriate regularization. Studying the structure of divergences, one gets access to the scaling equations and corresponding anomalous dimensions. In turn, it allows one to resum large logarithms of factorization scales and obtain scattering cross-sections in a wide kinematic range.

\begin{figure}[t]
\begin{center}
\includegraphics[width=0.45\textwidth]{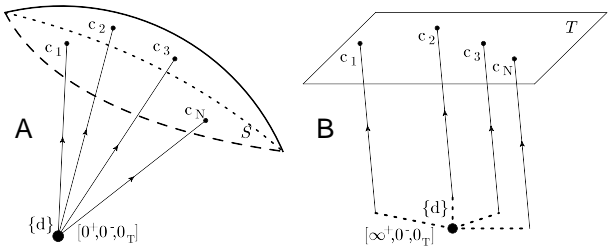}
\end{center}
\caption{Geometry of Wilson lines within matrix elements $\Pi^{\{cd\}}$ (A) and $\Xi^{\{cd\}}$ (B). The color indices attached to the end of Wilson lines are denoted by letters $c$ and $d$. The conformal transformation (\ref{def:conf_tranform}) maps the infinite sphere $S$ into the transverse plane $T$, and the origin to the lightlike infinity (black points).}\label{fig1}
\end{figure}

Considering the processes where several partons participate in \textit{single} hard interaction one often deals with SFs of the form
\begin{eqnarray}\label{def:JP_SF}
\mathcal{S}^{\{ad\}}(\{v\})=\sum_X w_X \Pi^{\dagger\{ac\}}_X(\{v\})\Pi^{\{cd\}}_X(\{v\}),
\end{eqnarray}
where $X$ denote the complete set of states, and $\{v\}=v_1,...,v_N$ are vectors pointing along the momenta of scattering partons. The function $\Pi$ is given by
\begin{eqnarray}\label{def:PI}
\Pi^{\{cd\}}_X(\{v\})=\langle X|T\[\Phi^{c_1d_1}_{v_1}(0)...\Phi^{c_Nd_N}_{v_N}(0)\] |0\rangle.
\end{eqnarray}
where $\Phi_v(x)$  is a half-infinite Wilson line rooted at $x$ and pointing in the direction $v$
\begin{eqnarray}\label{def:Phi}
\mathbf{\Phi}_v(x)=P\exp\(ig \int_0^\infty d\sigma v^\mu A_\mu^A(v \sigma+x)\mathbf{T}^A\),
\end{eqnarray}
with $\mathbf{T}^A$ being the generator of the gauge group. The pictorial representation of $\mathbf{\Pi}$ is shown in fig.\ref{fig1}A. Here and later the bold font denotes the objects with matrix color structure, that in eqns. (\ref{def:JP_SF},\ref{def:PI}) is represented by indices $a,c,d$.  The weight function $w_X$ strongly depends on the type of the factorization theorem and the process under consideration. SFs similar to (\ref{def:JP_SF}) are very generic and arise in various  applications. The most popular examples are: the description of multijet production and event shapes (see e.g. \cite{Ellis:2009wj,Bauer:2008jx,Ellis:2010rwa}), the hard-collinear factorization and Sudakov resummation for several partons (see e.g. \cite{Contopanagos:1996nh,Kidonakis:1998nf,Feige:2014wja,Bauer:2001yt,Bonciani:2003nt,Becher:2009cu,Becher:2009qa}), and threshold resummation  (see e.g.\cite{Kidonakis:1998bk,Laenen:1998qw,Becher:2007ty}). In this letter we discuss only configurations with lightlike vectors $v$ ($v_i^2=0$) which correspond to scattering of massless or high-energetic partons.

A different configuration appears in the multiparton scattering \cite{Diehl:2011yj,Manohar:2012jr,Diehl:2015bca,Vladimirov:2016qkd}. In this case partons scatter \textit{pairwise}, and SF reads
\begin{eqnarray}\label{def:MPS_SF}
&&\Sigma^{\{ad\}}(\{b\})=
\\\nn&&
\langle 0|T\big[\big(\Phi_{-\bar n}\Phi_{-n}^\dagger\big)^{a_1d_1}(b_{1})...\big(\Phi_{-\bar n}\Phi_{-n}^\dagger\big)^{a_Nd_N}(b_{N})\big]|0\rangle,
\end{eqnarray}
where $n$ and $\bar n$ denote a pair of lightlike vectors $n^2=\bar n^2=0$ and $b$ are vectors in the transverse plane $(nb_i)=(\bar n b_i)=0$. The configuration (\ref{def:MPS_SF}) is  typical for Drell-Yan-like processes, where all partons belong to initial hadrons. Such processes represent a part of background interactions at high energies. All intervals within the operator (\ref{def:MPS_SF}) are space- or lightlike. It allows one to rewrite $\Sigma$ in the form similar to (\ref{def:JP_SF})
\begin{eqnarray}\label{def:MPS_SF=XI^2}
\mathbf{\Sigma}(\{b\})=\sum_X \tilde w_X \mathbf{\Xi}^\dagger_{\bar n,X}(\{b\})\mathbf{\Xi}_{n,X}(\{b\}),
\end{eqnarray}
where $\tilde w_X$ is a unity weight and
\begin{eqnarray}\label{def:XI}
\Xi^{\{cd\}}_{n,X}(\{b\})&=&\langle X|T\[\Phi^{\dagger c_1d_1}_{-n}(b_{1})...\Phi^{\dagger c_Nd_N}_{-n}(b_{N})\] |0\rangle.
\end{eqnarray}
The pictorial representation of $\mathbf{\Xi}$ is shown in fig.\ref{fig1}B. SFs $\mathbf{\Sigma}$ are not studied in such details as SFs $\pmb{\mathcal{S}}$, mostly because the subject is relatively new. 

The SFs $\pmb{\mathcal{S}}$ and $\mathbf{\Sigma}$ as written in (\ref{def:JP_SF}) and (\ref{def:MPS_SF}) are color multimatrices. Generally, constituent Wilson lines $\Phi$ are of different color representations.
%, i.e. indices $a,b,c$ can belong to fundamental, anti-fundamental, adjoint (which in QCD corresponds to quark, anti-quark, gluon) or any other gauge group representation. 
Under the gauge rotation the sets of indices $\{a\}$ or $\{d\}$ are transformed by gauge transformation matrices at the same points \footnote{Additional transverse Wilson links can be added to the definition of (\ref{def:XI}) to enforce the gauge invariance \cite{GarciaEchevarria:2011md}. It is not necessary if zero boundary condition at lightlike infinities is used, e.g. in covariant gauges.}. Therefore, the singlet elements of SFs are gauge invariant. Naturally, only such combinations contribute into the factorization formulas.

At $N=2$ SFs $\pmb{\mathcal{S}}$ and $\mathbf{\Sigma}$ are given by the same configuration of Wilson lines. It appears within the transverse momentum dependent (TMD) factorization theorem which describes, e.g. unintegrated Drell-Yan process \cite{Bauer:2001yt,Collins:2011zzd}. In our notation this important case reads
\begin{eqnarray}\label{def:TMD_SF}
S(b)=\frac{1}{N_{rep}}\Tr\, \pmb{\mathcal{S}}(n,\bar n)=\frac{1}{N_{rep}}\Tr \mathbf{\Sigma}(b,0),
\end{eqnarray}
where $N_{rep}$ is the dimension of Wilson lines representation. Here for $\mathcal{S}$ we take the weight $w_X=e^{i\hat Pb}$ with $\hat{P}$ being the shift operator, and use the Lorentz boost to the frame where $\{v_1,v_2\}=\{n,\bar n\}$. TMD SF has been studied in detail in various regularizations, see e.g. recent two-loop evaluations \cite{Echevarria:2015byo,Luebbert:2016itl,Li:2016axz}.

Generally SFs have ultraviolet (UV) and infrared (IR) divergences (in some cases the latter are regularized by the weight function $w_X$). While UV divergences are removed by the renormalization procedure, the IR divergences are an inherent part of SFs. For some configurations the IR divergences related to different partons can be separated. In these cases universal process-independent parton substructures can be defined. In particular,the separation of rapidity divergences for the TMD SF \cite{Collins:2011zzd,Chiu:2012ir} results in well-defined TMD parton distributions \cite{Collins:2011zzd,Echevarria:2012js,Echevarria:2016scs}. The similar statement for the double-parton scattering ($\mathbf{\Sigma}$ at $N=4$) has been recently observed at two-loop order \cite{Vladimirov:2016qkd}. In this Letter we prove that rapidity divergences can be removed by the renormalization procedure for every $\mathbf{\Xi}$. It is equivalent to the statement that rapidity divergences for SFs $\mathbf{\Sigma}$ are factorizable.

\textbf{Correspondence between $\Xi$ and $\Pi$}.
Individually $\mathbf{\Pi}$ and $\mathbf{\Xi}$ are not well-defined. First, they are not gauge invariant even if color indices form singlets. Indeed, under the gauge transformation  indices $\{c\}$ of $\Pi^{\{cd\}}$ couple to gauge matrices at different points of infinite lightlike sphere, and indices $\{c\}$ of $\Xi^{\{cd\}}$ couple to gauge matrices at different points in the transverse plane. Second, $\mathbf{\Pi}$ and $\mathbf{\Xi}$ have different set of divergences than SFs. The divergences that appear due to the interaction between parts of SFs are not present. Instead, a different set of divergences appears. One example is the UV divergences of $\mathbf{\Sigma}$ that arise due to the gluon exchanges between $\mathbf{\Phi}^\dagger_{-n}$ and $\mathbf{\Phi}_{-\bar n}$ in the vicinity of their contact point. These divergences are absent in the matrix element $\mathbf{\Pi}$, which however has end-point divergences due to non-analytical behavior at the end of a single $\mathbf{\Phi}^\dagger_{-n}$. Another example is mass divergences. Mass-divergences cancel in the sum of all diagrams for a SF, but remain uncanceled within matrix elements $\mathbf{\Pi}$ and $\mathbf{\Xi}$. In total, $\mathbf{\Pi}$ and $\mathbf{\Xi}$ are strongly dependent on the gauge fixation condition and the regularization scheme. %For example, end-point divergences would disappear if one fixes the gauge such that fields vanish at the boundary where end-points located. Within the set of such gauges $\mathbf{\Pi}$ and $\mathbf{\Xi}$ are invariant. For example, $\mathbf{\Pi}$ is often considered as it stands \cite{Feige:2014wja,Contopanagos:1996nh,Kidonakis:1998nf,Bonciani:2003nt,Becher:2009qa} within covariant gauges (where $A_\mu(x)=0$ at $x\to \infty$). 
%In this case it is independent of the choice of gauge-fixing parameter. 
In the following, we assume that $\mathbf{\Pi}$ and $\mathbf{\Xi}$ are considered within appropriate gauge and successful IR regularization, which we call a calculation scheme for brevity.

Despite the strong general dependence on the calculation scheme, the matrix elements $\mathbf{\Pi}$ and $\mathbf{\Xi}$ have substructures that are insensitive to it
%The matrix elements $\mathbf{\Pi}$ and $\mathbf{\Xi}$ have substructures that are insensitive to the problems of general ill-definition. 
These are UV divergences for $\mathbf{\Pi}$ and rapidity divergences for $\mathbf{\Xi}$. The UV divergences of $\mathbf{\Pi}$ are removed by a renormalization matrix \cite{Brandt:1981kf,Dotsenko:1979wb}
\begin{eqnarray}\label{def:Z}
\mathbf{\Pi}^{bare}_X(\{v\})=\mathbf{\Pi}_X(\{v\})\mathbf{Z}(\{v\}).
\end{eqnarray}
The matrix $\mathbf{Z}$ is independent of $X$ and of the calculation scheme. Therefore, the renormalization scale dependence is also calculation scheme independent
\begin{eqnarray}\label{def:sAD}
\mu^2 \frac{d}{d\mu^2}\mathbf{\Pi}_X(\{v\})=\mathbf{\Pi}_X(\{v\})\pmb{\gamma}_s(\{v\}),
\end{eqnarray}
where  $\pmb{\gamma}_s(\{v\})$ is the soft anomalous dimension (sAD). The sAD is an essential part of factorization, and nowadays known up to three-loop order \cite{Almelid:2015jia}.

The matrix element $\mathbf{\Pi}_{X=0}$ can be transformed to the matrix element $\mathbf{\Xi}_{X=0}$ by the conformal-stereographic projection \cite{Cornalba:2007fs}
%To demonstrate it let us consider a single Wilson line entering the definition of (\ref{def:PI})
%\begin{eqnarray}\label{conf:Phi=[a,b]}
%\Phi^{ab}_v(0)=\!\!\!\lim_{\substack{\alpha \to +0\\\beta\to+\infty}}\[\{\alpha v^+,\alpha v^-,\alpha v_T\},\{\beta v^+,\beta v^-,\beta v_T\}\]^{ab}\!\!\!,
%\end{eqnarray}
%where we denote a Wilson line from point $a$ to $b$ as $[a,b]$, and use popular notation for the components of a vector $v^+=(nv)$, $v^-=(\bar nv)$ and %$(nv_T)=(\bar n v_T)=0$. Next, we consider the following conformal transformation 
\begin{eqnarray}\label{def:conf_tranform}
\mathcal{C}:\{x^+,x^-,x_T\}\to \Big\{-\frac{1}{2x^+},x^--\frac{x_T^2}{2x^+},\frac{x_T}{\sqrt{2}x^+}\Big\},
\end{eqnarray}
where we use common notation for the components of a vector $v^+=(nv)$, $v^-=(\bar nv)$ and $(nv_T)=(\bar n v_T)=0$.
Applying this transformation to the $\Phi^{ab}_v(0)$ we obtain
\begin{eqnarray}
\mathcal{C}\Phi^{cd}_v(0)&=&
%\lim_{\substack{\alpha \to +0\\\beta\to+\infty}}\[\Big\{\frac{-1}{\sqrt{2}\alpha},0,v_T\Big\},
%\Big\{\frac{-1}{\sqrt{2}\beta},0,v_T\Big\}\]^{ab}
%\\
%&=&
\Phi^{\dagger\,cd}_{-n}\(\frac{v_T}{\sqrt{2}v^+}\).
\end{eqnarray}
Therefore, under the transformation $\mathcal{C}$ the matrix element $\mathbf{\Pi}$ transforms as (see also fig.\ref{fig1})
\begin{eqnarray}\label{PI->XI}
\mathbf{\Pi}_{X=0}(\{v\})\to \mathbf{\Xi}_{X=0}\(\{b\}\).
\end{eqnarray}
with $\{b\}=\{v_T/\sqrt{2}v^+\}$. 
In the following we discuss the consequences of (\ref{PI->XI}) for the structure of divergences.

Transforming $\mathbf{\Pi}$ to $\mathbf{\Xi}$ by the conformal transformation $\mathcal{C}$ one also changes the classification of divergences.
% Changing the geometry of Wilson lines the conformal transformation $\mathcal{C}$ also changes the structure of divergences. 
So, it can be traced that the UV divergences of $\mathbf{\Pi}$ are mapped into the rapidity divergences of $\mathbf{\Xi}$. Therefore, in a \textit{conformal} field theory the UV finite expression (\ref{def:Z}) turns into the rapidity-divergence-free expression for $\mathbf{\Xi}$
\begin{eqnarray}\label{def:R}
\mathbf{\Xi}^{div}_{n,X}(\{b\})=\mathbf{\Xi}_{n,X}(\{b\})\mathbf{R}(\{b\}),
\end{eqnarray}
where the matrix $\mathbf{R}$ is obtained from the matrix $\mathbf{Z}$ by some transformation of regularizations. The renormalization of rapidity divergences introduces the scale parameter $\zeta$. The dependence on it is given by the rapidity anomalous dimension (rAD)
\begin{eqnarray}
\zeta \frac{d}{d\zeta}\mathbf{\Xi}_{n,X}(\{b\})=2\mathbf{\Xi}_{n,X}(\{b\})\mathbf{D}(\{b\}),
\end{eqnarray}
where the factor $2$ is put to match the definition of TMD rAD. This relation shows that rapidity divergences of $\mathbf{\Xi}$ are independent of $X$ (which also follows from the general consideration \cite{Erdogan:2014gha}), and rAD is independent of the calculation scheme (which is known in the TMD case \cite{Collins:2012uy,Echevarria:2016scs,Li:2016axz,Luebbert:2016itl}). The renormalization of rapidity divergences in the form (\ref{def:R}) was used for the case of TMD SF in many papers (see e.g.\cite{Echevarria:2016scs,Chiu:2012ir}). In the following we argue that (\ref{def:R}) is also valid for pQCD, where conformal symmetry is violated by quantum corrections.

The matrices $\mathbf{Z}$ and $\mathbf{R}$ are the same matrices if calculated in some proper regularizations. Therefore, within \textit{conformal} field theory we have the relation
\begin{eqnarray}\label{gamma=D}
\pmb{\gamma}_s(\{v\})=2\mathbf{D}(\{b\}).
\end{eqnarray}
This relation for TMD kinematics has been recently confirmed for $\mathcal{N}=4$ SYM theory at three-loop order in \cite{Li:2016ctv}.

Note, that the transformation (\ref{def:conf_tranform}) is also used to relate the Balitsky-Kovchegov (BK) equation with Banfi-Marchesini-Smye (BMS) equations \cite{Hatta:2008st,Avsar:2009yb}. In this case, the UV singularity is mapped onto collinear one and results in the same non-linear evolution equation.

\textbf{Relation between rAD and sAD in pQCD}. 
In QCD the conformal invariance and hence the relation (\ref{gamma=D}) are violated by quantum corrections. Nonetheless, the violating terms can be found by studying eqn.(\ref{gamma=D}) at critical coupling where the conformal invariance of QCD is restored \cite{Banks:1981nn}. The value of critical coupling $a_s^*$ is defined by the zero of QCD $\beta$-function. In the dimension regularization (with $D=4-2\epsilon$) and $\mathrm{MS}$-scheme the $\beta$-function is
\begin{eqnarray}
\beta(g)=g(-\epsilon-a_s \beta_0-a_s^2\beta_1-...),
\end{eqnarray}
where $g$ is QCD coupling constant, $a_s=g^2/(4\pi)^2$ and the coefficients $\beta_n$ are well known. The equation $\beta(g^*)=0$ defines the value of $a_s^*(\epsilon)$. Equivalently one can find the number of space-time dimension $\epsilon^*$ at which QCD turns to the critical regime, $\epsilon^*=-a_s\beta_0-a_s^2\beta_1-...$\,. 

The essential statement of the approach is that in MS-like schemes UV anomalous dimensions are \textit{independent} on the choice of $\epsilon$ \cite{Braun:2013tva,Braun:2016qlg,Vasilev:2004yr}. In other words, sAD has full conformal symmetry and is the same for the physical QCD and the QCD at critical coupling. The rAD does depend on $\epsilon$. At critical coupling, the conformal invariance of QCD is restored and the relation (\ref{gamma=D}) holds. Therefore, in QCD we have
\begin{eqnarray}\label{QCD:gamma=D}
\pmb{\gamma}_s(\{v\})=2\mathbf{D}(\{b\},\epsilon^*).
\end{eqnarray}
This relation presents our main result. It connects at all orders of perturbation theory the sAD with the rAD evaluated in the particular (critical) number of dimensions.

Let us test the relation (\ref{QCD:gamma=D}) in the simplest case at $N=2$. In this case the matrix structure is reduced to a single entry (\ref{def:TMD_SF}). The sAD has the form \cite{Kidonakis:1998nf,Aybat:2006wq,Gardi:2009qi}
\begin{eqnarray}
\gamma_s(v_1,v_2)=\Gamma_{\text{cusp}}(a_s)\ln\(\frac{v_{12}\mu^2}{\nu^2}\)-\tilde \gamma_s(a_s),
\end{eqnarray}
where $\Gamma_{\text{cusp}}$ is the lightlike cusp-anomalous dimension, $v_{ij}=(v_i-v_j)^2$, and $\nu$ is some overall IR scale. The coefficients of perturbative expansions $\Gamma_{\text{cusp}}=4 C_f a_s+\Gamma_1 a_s^2+...$ and $\tilde \gamma_s(a_s)=0\cdot a_s+\tilde \gamma_{1s}a_s^2+...$  are known up to order $a_s^3$ \cite{Moch:2004pa,Moch:2005tm} (here $C_f$ is the quadratic Casimir eigenvalue for the representation $f$, dictated by the representation of Wilson lines). At $N=2$ the rAD is reduced to TMD rAD \cite{Collins:2011zzd,Echevarria:2015byo,Luebbert:2016itl,Echevarria:2012js,Echevarria:2016scs,Chiu:2012ir}
\begin{eqnarray}
\tr\mathbf{D}(b_1,b_2)=N_{rep}\mathcal{D}(L,a_s),
\end{eqnarray}
where $L=\ln(\mu^2 b_{12}e^{2\gamma_E}/4)$ with $b_{ij}=(b_i-b_j)^2$.
% and function $\mathcal{D}$ is known up to $a_s^3$ order \cite{Li:2016ctv}. 
The dependence of rAD on the renormalization scale is given by 
\begin{eqnarray}\label{QCD:RGE_D}
\mu^2 \frac{d}{d\mu^2}\mathcal{D}(L,a_s(\mu))=\frac{\Gamma_{\text{cusp}}(a_s)}{2},
\end{eqnarray}
which fixes the logarithmic part of $\mathcal{D}$ in terms of $\Gamma_{\text{cusp}}$ and $\beta$-function (see e.g. collection of formulas in \cite{Echevarria:2016scs}). 

Applying the perturbative expansion to eqn.(\ref{QCD:gamma=D}) one can obtain higher order terms of $\mathcal{D}$ from the lower order terms and $\gamma_s$. To obtain $a_s^3$-term one needs to consider rAD at $a_s^2$-order at arbitrary $\epsilon$. It can be found from the calculation made in  \cite{Echevarria:2015byo}
\begin{eqnarray}\nn
&&\mathcal{D}(L,a_s,\epsilon)=-2 a_s C_f \(B^{\epsilon} \Gamma(-\epsilon)+\frac{1}{\epsilon}\)+2 C_fa_s^2\Big\{
\\\nn 
&&\quad B^{2\epsilon}\Gamma^2(-\epsilon)\Big[C_A\big(2\psi(-2\epsilon)-2\psi(-\epsilon)+\psi(\epsilon)+\gamma_E\big)
\\\nn&&\quad+\frac{1-\epsilon}{(1-2\epsilon)(3-2\epsilon)}\(\frac{3(4-3\epsilon)}{2\epsilon}C_A-N_f\)\Big]
\\&&\quad+B^\epsilon \frac{\Gamma(-\epsilon)}{\epsilon}\beta_0+\frac{\beta_0}{2\epsilon^2}-\frac{\Gamma_1}{2\epsilon}\Big\}+\mathcal{O}(a_s^3),
\end{eqnarray}
where $B=e^{L}$, and $N_f$ is number of fermions. Under the conformal transformation (\ref{def:conf_tranform}) the variable $v_{ij}$ turns to $b_{ij}$ with unknown prefactor, which cannot be fixed due to lightlike origin of vectors $v$. This prefactor can be absorbed into the variable $\nu$, which we find by matching the leading order expressions $\nu=2e^{-\gamma_E}$.  Considering the perturbative expansion at critical $\epsilon^*$ we obtain
\begin{eqnarray}\label{QCD:gamma=D+deltaD}
\Gamma_{\text{cusp}}(a_s)L-\tilde \gamma_s(a_s)=2\mathcal{D}(L,a_s)+2\delta \mathcal{D}(L,a_s),
\end{eqnarray}
where 
\begin{eqnarray}\nn
\delta\mathcal{D}(L,a_s)=-a_s^2C_f\beta_0(L^2+\zeta_2)+a_s^3C_f\Big[\qquad
\\\nn
-\frac{2\beta_0^2}{3}L^3 -\(\frac{\beta_0\Gamma_1}{2}+\beta_1\)L^2+\beta_0(\tilde \gamma_{1s}-2\beta_0\zeta_2)L
\\\nn -\beta_0\Gamma_1\frac{\zeta_2}{4}
-\zeta_2\beta_1+\frac{2\beta_0^2}{3}\(\zeta_3-\frac{82}{9}\)
\\+26\beta_0C_A\(\zeta_4-\frac{8}{27}\)\Big]+\mathcal{O}(a_s^4).\label{QCD:deltaD}
\end{eqnarray}
Comparing the perturbative orders of eqn.(\ref{QCD:gamma=D+deltaD}) we express TMD rAD in terms of sAD (or vise-versa) up to order $a_s^3$. The finite part of $\mathcal{D}$ is
\begin{eqnarray}\nn
\mathcal{D}(0,a_s)\!=\!a_s^2\(-\!\frac{\tilde \gamma_{1s}}{2}\!+\!C_f\beta_0\zeta_2\)\!+\!a_s^3\(-\!\frac{\tilde \gamma_{2s}}{2}\!-\!X\)\!+\!\mathcal{O}(a_s^4),
\end{eqnarray}
where $X$ is given by the last two lines of (\ref{QCD:deltaD}). This expression coincides with the result of the direct calculation \cite{Li:2016ctv}.  The logarithmic part of $\mathcal{D}$ is dictated by eqn.(\ref{QCD:RGE_D}) and exactly reproduced in (\ref{QCD:gamma=D+deltaD}). This example gives a non-trivial confirmation of correspondence presented in this Letter.

Using (\ref{QCD:gamma=D}) we can also derive the rAD for a general multiparton scattering configuration at $a_s^3$-order. Its color structure has the form
\begin{eqnarray}\label{QCD:MPS_D}
\mathbf{D}(\{b\})=\sum_{1\leqslant i<j}^N\mathbf{T}^A_i \mathbf{T}^A_j \mathcal{D}(L_{ij},a_s)\qquad\qquad\qquad
\\\nn
+f^{AB\alpha}f^{\alpha CD}\Big[\sum_{i,j,k,l}\mathbf{T}_i^A\mathbf{T}_j^B\mathbf{T}_k^C\mathbf{T}_l^D\tilde{\mathcal{F}}(b_i,b_j,b_k,b_l)
\\\nn+\sum_{i=1}^N\sum_{\substack{1\leqslant j<k\\ i\neq j,k}}^N\{\mathbf{T}_i^A,\mathbf{T}_i^D\}\mathbf{T}_j^B\mathbf{T}_k^C \tilde C(b_i,b_j,b_k)\Big]+...
\end{eqnarray}
where indices $i,j,k,l$ enumerate Wilson lines in $\mathbf{\Xi}$, the summation indices in the second line are all different, $f^{ABC}$ is the gauge-group structure constant and dots denote the color structures that appear at higher perturbative orders. The functions $\tilde{\mathcal{F}}$ and $\tilde C$ are of order $a_s^3$. We have obtained the expression (\ref{QCD:MPS_D}) by resolving the color algebra of $\mathbf{\Sigma}$ in terms of the generating function for web-diagrams \cite{Vladimirov:2014wga,Vladimirov:2015fea}. Note the absence of color structures like $f^{ABC}\mathbf{T}^A_i\mathbf{T}^B_j\mathbf{T}^C_k$ or $d^{ABC}\mathbf{T}^A_i\mathbf{T}^B_j\mathbf{T}^C_k$, which in principle could appear at this order. In the limit $N=2$ all the color structures except the leading one are zero, therefore, the first line represents TMD rAD (in sAD terminology it is called the dipole contribution).   Eqn.(\ref{QCD:MPS_D}) agrees with the explicit two-loop calculation of $\mathbf{D}$ \cite{Vladimirov:2016qkd}.

The non-dipole contribution to sAD evaluated at $a_s^3$-order in \cite{Almelid:2015jia} has the same color structure as found by us for rAD (\ref{QCD:MPS_D}), which grants an additional check for relation (\ref{QCD:gamma=D}). Due to the fact that functions $\tilde{\mathcal{F}}$ and $\tilde C$ are of $a_s^3$ order, they are not affected by conformal-symmetry violating corrections at this order. Therefore, comparing with \cite{Almelid:2015jia} we obtain
\begin{eqnarray}
&&\tilde{\mathcal{F}}(b_i,b_j,b_k,b_l)=a_s^3\mathcal{F}(\rho_{ikjl},\rho_{iljk})+\mathcal{O}(a_s^4),
\\
&&\tilde C(b_i,b_j,b_k)=a_s^3\(-\zeta_2\zeta_3-\zeta_5/2\)+\mathcal{O}(a_s^4),
\end{eqnarray}
where $\rho_{ijkl}=b_{ij}b_{kl}/b_{ik}b_{jl}$ is the conformal ratio. The explicit form of function $\mathcal{F}$ can be found in \cite{Almelid:2015jia}.

\textbf{Conclusion \& discussion}.
We have shown that UV divergence of the "multi-cusp" configuration of lightlike Wilson lines is related by the conformal transformation to the rapidity divergence of set of parallel lightlike half-infinite Wilson lines. This correspondence is exact in any conformal field theory and violated by $\beta$-function corrections in QCD. The main consequence of the correspondence is that rapidity divergences can be removed from a (matrix) SF (\ref{def:MPS_SF}) by a singular (matrix) factor, analogous to UV renormalization factor. This statement holds at arbitrary perturbative order in conformal field theory \textit{and QCD} (at least in MS-like schemes) because the $\beta$-function corrections modify only numerical values of factors but not the color or divergence structure. Therefore, we have proven the renormalization of rapidity divergences for configurations of Wilson lines of type (\ref{def:MPS_SF}). The proof of rapidity renormalization procedure supplements many factorization theorems and allows rigorous definition of corresponding parton distribution, such as TMD parton distribution and double-parton distributions. In the particular case of TMD kinematics the rapidity renormalization is known and has been discussed in many papers, e.g. \cite{Collins:2011zzd,Echevarria:2015byo,Li:2016axz,Luebbert:2016itl,Echevarria:2012js,Echevarria:2016scs,Chiu:2012ir}. For the case of double-parton scattering it has been demonstrated at two-loop order in \cite{Vladimirov:2016qkd}.

The equivalence of rapidity and UV renormalization factors leads to the equality of rapidity and soft anomalous dimensions in a conformal field theory, which was also observed in direct calculations \cite{Li:2016ctv}. In QCD the terms violating this equality can be found by considering QCD at the critical coupling. We have derived the violating terms for TMD kinematics up to $a_s^3$ order and confirmed the result for the recent three-loop evaluation of rAD made in \cite{Li:2016ctv}. Also, we  have obtained the general matrix-valued rAD for a multiparton scattering at order $a_s^3$. The obtained expressions at order $a_s^2$ coincide with results presented in \cite{Vladimirov:2016qkd}. The color structure of the general matrix-valued rAD repeats those of sAD, what we have checked explicitly within the generating function approach \cite{Vladimirov:2014wga,Vladimirov:2015fea}. The leading color-quadrupole contribution is obtained from the corresponding terms of sAD  \cite{Almelid:2015jia}, and is a new result.

The presented study makes a bridge between seemingly very different kinematic regimes, namely the multiparticle production and the multiparton scattering.
 The exact relation between sAD and rAD is an immediate result, and probably much more would come with a deeper study.

\acknowledgments The author is grateful to V.Braun, A.Manashov and I.Scimemi for multiple discussions, and also to D.Neill for the introduction into BK-to-BMS relation which initiated this research.

\end{document}